\pdfoutput=1

\documentclass[11pt,twoside,a4paper,cmspaper,final,collab]{cms-tdr}
\def\svnVersion{175953}\def\cmsCernNo{2012-312}\def\cmsCernDate{\today}\def\cmsMessage{Submitted to the Journal of High Energy Physics}

\begin{document}\cmsNoteHeader{EXO-12-014}

\hyphenation{had-ron-i-za-tion}
\hyphenation{cal-or-i-me-ter}
\hyphenation{de-vices}

\RCS$Revision: 163659 $
\RCS$HeadURL: svn+ssh://svn.cern.ch/reps/tdr2/papers/EXO-12-014/trunk/EXO-12-014.tex $
\RCS$Id: EXO-12-014.tex 163659 2013-01-08 00:02:27Z cleonido $
\newlength\cmsFigWidth
\ifthenelse{\boolean{cms@external}}{\setlength\cmsFigWidth{0.85\columnwidth}}{\setlength\cmsFigWidth{0.4\textwidth}}
\ifthenelse{\boolean{cms@external}}{\providecommand{\cmsLeft}{top}}{\providecommand{\cmsLeft}{left}}
\ifthenelse{\boolean{cms@external}}{\providecommand{\cmsRight}{bottom}}{\providecommand{\cmsRight}{right}}
\cmsNoteHeader{EXO-12-014} 
\providecommand\Wprime{\PWpr\xspace}

\newcommand{\rhoT}{\ensuremath{\rho_\mathrm{TC}}\xspace}
\newcommand{\aT}{\ensuremath{a_\mathrm{TC}}\xspace}
\newcommand{\omegaT}{\ensuremath{\omega_\mathrm{TC}}\xspace}
\newcommand{\piT}{\ensuremath{\pi_\mathrm{TC}}\xspace}
\newcommand{\wprime}{\ensuremath{W^{\prime}}\xspace}
\providecommand{\GKK}{\ensuremath{\mathrm{G}_\mathrm{KK}}\xspace}
\newcommand{\vpt}{\ensuremath{p^{V}_\mathrm{T}}\xspace}
\newcommand{\zpt}{\ensuremath{p^{Z}_\mathrm{T}}\xspace}
\newcommand{\kmpl}{\ensuremath{k/\overline{M}_\text{Pl}}\xspace}
\newcommand{\gtm}{\ensuremath{M_\mathrm{T}}\xspace}
\newcommand{\gravitonmass}{\ensuremath{M_\mathrm{G}}\xspace}
\newcommand{\jetmass}{\ensuremath{M_j}\xspace}

\newcommand{\usedLumiVZ}{5.0\fbinv\xspace}
\newcommand{\wprimeLimit}{XXX\GeV}
\newcommand{\rsLimit}{YYY\GeV}

\title{Search for exotic resonances decaying into $\PW\cPZ$/$\cPZ\cPZ$ in $\Pp\Pp$ collisions at $\sqrt{s} = 7\TeV$}

\date{\today}

\abstract{A search for new exotic particles decaying to the V$\cPZ$
  final state is performed, where V is either a $\PW $ or a $\cPZ$ boson
  decaying into two overlapping jets and the $\cPZ$ decays
  into a pair of electrons, muons or neutrinos.
 The analysis uses a data sample of pp collisions corresponding to an integrated luminosity of
 5\fbinv collected by the CMS experiment at the LHC at $\sqrt{s}=7$\TeV in 2011.
  No significant excess is observed
  in the mass distribution of the V$\cPZ$ candidates compared with the
  background expectation from standard model processes. Model-dependent upper limits
  at the 95\% confidence level are set on the product of the cross section
  times the branching fraction of hypothetical
  particles decaying to the V$\cPZ$ final state as a function of mass.
  Sequential standard model $\PWpr$ bosons with masses
  between 700 and 940\GeV are excluded. In the Randall--Sundrum model
  for graviton resonances with
  a coupling parameter of 0.05, masses
  between 750 and 880\GeV are also excluded.}

\hypersetup{%
pdfauthor={CMS Collaboration},%
pdftitle={Search for exotic resonances decaying into WZ/ZZ in pp collisions at sqrt(s) = 7 TeV},%
pdfsubject={CMS},%
pdfkeywords={CMS, physics, Wprime, Randall-Sundrum, RS, New Physics}}

\maketitle 
\section{Introduction}\label{intro}
In many extensions of the standard model (SM) of particle
physics~\cite{Glashow:1961tr,Weinberg:1967tq,Salam} the
spontaneous breaking of the electroweak (EW)
symmetry~\cite{Englert:1964et,Higgs:1964ia,Higgs:1964pj,Guralnik:1964eu,Higgs:1966ev,Kibble:1967sv}
is associated with new strong dynamics appearing at the\TeV scale. For
instance, the origin of the new dynamics may be due to new
interactions~\cite{Weinberg:orig,Weinberg:1979bn,Susskind:1978ms}, compact
extra
dimensions~\cite{Randall:1999ee,Randall:1999vf}, or a composite
Higgs boson~\cite{Kaplan:1983fs,Agashe:2004rs}. In such scenarios
the SM is an effective low-energy theory, valid for energies
smaller than a new-physics scale $\Lambda$. In these theories, one
expects the existence of new resonances coupling to pairs of vector
bosons ($\cPZ\cPZ$, $\PW\cPZ$, and $\PW \PW$).
A minimal
ultraviolet completion of this effective theory for
composite models is described in Ref.~\cite{Contino:2011np}.
Other examples
include Randall--Sundrum (RS) gravitons \GKK~\cite{Randall:1999ee,
  Randall:1999vf} coupled to $\cPZ\cPZ$ and $\PW \PW $, or
technimesons~\cite{TC1, TC2} coupled to $\PW \cPZ$. Limits from previous
searches and from indirect bounds (\eg in the EW sector and flavor physics)
place the masses of these proposed RS resonances at or above the\TeV scale
\cite{pomarol,   huber, Davoudiasl:2000wi,
  Collaboration:2011dca,Chatrchyan:2011fq,Abazov:2010dj,Aaltonen:2010cf}.
These scenarios could be tested at the Large Hadron Collider (LHC), as long as
$\Lambda \sim \mathcal{O}(\TeVns{})$, as suggested by the EW symmetry breaking scale.
This analysis is sensitive to searches for resonances starting at 700\GeV and above. However, there are other theories that predict light resonances
(\eg low-scale technicolor)~\cite{TC1, TC2}.

In this Letter we present a search for heavy resonances decaying to
WZ and ZZ final states,
with one boson being a $\cPZ$ decaying
to leptons, namely $\cPZ \to \ell^+ \ell^-$
($\ell = \mu,\ \Pe$) or $\cPgn\cPagn$, and the second boson decays
to hadrons, V (V=$\PW,\cPZ) \to \cPq \cPaq$.
For heavy resonances the decay of each V produces a highly boosted system
in which the two fermions are
emitted within a small opening angle in the laboratory frame.
The hadronization of the V$ \to \cPq\cPaq$
quarks would then produce two partially overlapping jets reconstructed as a
single jet with mass close to the V mass,
a topology very different from that of a typical quark or gluon jet. Monte Carlo (MC)
simulations suggest
that more than ${\sim}70\%$ of the decays would produce a merged-jet topology for
resonances heavier than ${\sim}800$\GeV. This feature is exploited in a
VZ final state, to
discriminate a possible signal from the SM background (mainly coming
from $\cPZ$+jets events).

Thus, in this study we
consider three final states: one heavy jet and
either $\cPZ \to \Pep\Pem$, $\cPZ \to \mu^+\mu^-$, or $\cPZ \to  \MET$, where \MET is the characteristic signature
of neutrino production.
We characterize the signal as a peak in the invariant mass of the VZ
system (transverse mass in the case of $\cPZ \to \cPgn\cPagn$ decays). The
search is performed with a data sample of pp collisions corresponding to an integrated luminosity of 5.0\fbinv collected by the Compact
Muon Solenoid (CMS) detector at the LHC at $\sqrt{s}=7$\TeV in 2011.

Results are presented in terms of two benchmark scenarios:
i) the Sequential Standard Model (SSM) in which a new gauge boson \Wprime with the same
couplings as the SM W boson decays to a WZ pair;
ii) a RS graviton, \GKK, decaying to ZZ. In both scenarios we search for resonances heavier
than 700\GeV, where the considered boosted topology becomes relevant.
The ratio of the 5-dimensional curvature to the reduced Planck
mass (\kmpl), which acts as the coupling constant in the RS model, is
typically used as the phenomenological parameter in RS graviton searches.
For the RS graviton scenario we consider values of the coupling parameter \kmpl up to 0.3.

Previous searches have been carried out in the context of both the SSM
\Wprime and RS graviton theoretical models. The most stringent limits have
been produced at the LHC by the ATLAS and CMS collaborations in a large
number of final states: $\Wprime \rightarrow \ell \nu$
\cite{Aad:2011yg,Chatrchyan:2012meb}, $\Wprime \rightarrow \cPqt\cPqb$
\cite{Aad:2012ej,:2012sc}, $\Wprime \rightarrow$ WZ $\rightarrow 3 \ell
\nu$ \cite{Aad:2012vs,Chatrchyan:2012kk}, and $\GKK \rightarrow \ell \ell$
\cite{Collaboration:2011dca,Chatrchyan:2012it}, $\GKK \rightarrow \gamma
\gamma$ \cite{ATLAS:2011ab,Chatrchyan:2011fq} and $\GKK \rightarrow$ ZZ $\rightarrow \ell \ell \, jj$ \cite{Collaboration:2012iua,CMS-PAS-EXO-11-102}.

\section{The CMS Detector} \label{sec:CMS}
Here a brief description of the CMS detector is given with an emphasis on the elements most relevant for this analysis.
A more detailed description can be found elsewhere~\cite{JINST}.
A cylindrical coordinate system about the beam axis
is used, in which the polar angle $\theta$ is
measured with respect to the counterclockwise beam direction and the
azimuthal angle, $\phi$, is measured in the $x$-$y$ plane in radians, where the $x$ axis
points towards the center of the LHC ring.
The quantity $\eta$ is the
pseudorapidity, defined as $\eta = -\ln [\tan (\theta /2)]$.
The layout comprises a superconducting solenoid providing a uniform magnetic field of 3.8\unit{T}.
The bore of the solenoid is instrumented with various particle detection systems.
The inner tracking system is composed of a pixel detector with three barrel layers at radii
between 4.4 and 10.2\unit{cm} and a silicon strip tracker with 10 barrel detection layers extending
outwards to a radius of 1.1\unit{m}. Each system is completed by two end-caps, extending
the acceptance up to $| \eta | < 2.5$.
A lead-tungstate crystal electromagnetic calorimeter with fine transverse ($\Delta \eta, \Delta \phi$)
granularity and a brass/scintillator hadronic calorimeter surround the tracking volume and cover the
region $ | \eta | < 3$. CMS also has extensive forward calorimetry.
The steel return yoke outside the solenoid is instrumented with gas-ionization detectors
which are used to identify muons in the range $ | \eta | < 2.4$.
The barrel region is covered by drift tube chambers and the end cap region by cathode strip chambers,
each complemented by resistive plate chambers.

\section{Collision Data and Monte Carlo Samples}\label{samples}

The preselection of the datasets for the
analysis is different for
the ``dilepton'' (VZ$ \rightarrow \cPq \cPaq\,\ell^+ \ell^-, \; \ell = \Pe, \mu$) and the
``\MET'' (VZ$ \rightarrow \cPq \cPaq \, \cPgn\cPagn$) channels. For the dilepton
channels,  we consider events that were recorded
with double-electron or single-muon triggers.
The trigger thresholds changed with time, as a consequence of the increasing peak
luminosity and the changes in running conditions. The tightest thresholds used in
the trigger (\ie 40\GeV for the single-muon trigger and 17\GeV for the dielectron trigger)
are looser than the corresponding offline analysis requirements.
Typical trigger efficiencies exceed 83\% (95\%) for the
electron (muon) triggers. For the \MET channel, we use triggers requiring
at least one calorimetric jet and missing transverse energy. These triggers have efficiencies
of more than 99\% for events
with a leading jet of transverse momentum $\pt > 160\GeV$ and $\MET > 300\GeV$
after offline reconstruction and corrections,
which allows resonances heavier than 1000\GeV to be probed with an efficiency above $20 \%$.
We use MC samples to study the signal and
background. We consider the SM background processes that could
contribute with two leptons and a (massive) jet in the final state. The
summary of
the signal samples is given in
Table~\ref{tab:sigSamples}, and the background samples, in Table~\ref{tab:bkgSamples}.
The \PYTHIA 6.424 \cite{Sjostrand:2006za} leading-order (LO) generator with
tune Z2 \cite{Chatrchyan:2011id} is used to generate the signal events
and simulate the parton showering,
with a full simulation of the detector based on \GEANTfour 9.4 package.
Mass-dependent $K$ factors are applied.
For the \GKK analysis, next-to-leading order (NLO) corrections are calculated using the
``two cutoff phase space slicing'' method~\cite{Kumar200945,Kumar200928}
in the diphoton final state.
For the \Wprime analysis, the next-to-next-to-leading order (NNLO) corrections are calculated with
\textsc{fewz}~\cite{Gavin:2010az} in the leptonic final state.
These $K$ factors are used for lack of better
(N)NLO calculations for the final states considered.
The background samples are
generated using the \MADGRAPH 5.1.1.0 matrix-element generator
\cite{madgraph,Herquet:2008zz}, while {\sc Pythia} is used for the
parton showering and hadronization,
with the same version and tuning as for signal samples.
The parton distribution function (PDF) used is {CTEQ6L1}~\cite{ref:CTEQ6L1}.
Jets are matched to partons using
the MLM scheme \cite{MLM}.

\begin{table}[htb]
  \topcaption{Signal Monte Carlo samples. The listed cross sections are \PYTHIA LO, per channel ($\Pep\Pem$ or $\mu^+\mu^-$ or $\cPgn\cPagn$). The
    notation $\cPgn\cPagn$ includes all three neutrino flavors. The $K$
    factors comprise NLO (NNLO) corrections for the \GKK~(\Wprime)
    samples. The \GKK samples are generated with $\kmpl = 0.05$.}
 \centering{
   \tabcolsep 3.5pt
  \begin{tabular}{|c|c|c|c|}
  \hline\hline
 Mass (\GeVns{}) & \multicolumn{2}{|c|}{Cross Section (pb)} & $K$ factor \\
\hline
{} &  {$\GKK \to \cPq\cPaq\,\ell^+ \ell^-$ ($\Pep\Pem$ or $\mu^+\mu^-$)} &
{$\GKK \to \cPq\cPaq \,\cPgn\cPagn$} & \\
\hline
 750  & 8.35$\times 10^{-3}$ & 5.03$\times 10^{-2}$ & 1.75 \\
 1000 & 1.52$\times 10^{-3}$ & 9.09$\times 10^{-3}$ & 1.78  \\
 1250 & 3.47$\times 10^{-4}$ & 2.16$\times 10^{-3}$ & 1.79 \\
 1500 & 8.83$\times 10^{-5}$ & 5.24$\times 10^{-4}$ & 1.78  \\
 1750 & 3.43$\times 10^{-5}$ & 2.04$\times 10^{-4}$ & 1.76  \\
 2000 & 7.05$\times 10^{-6}$ & 4.18$\times 10^{-5}$ & 1.76 \\
  \hline
{} &  { $\Wprime \to \cPq\cPaq\,\ell^+ \ell^-$ ($\Pep\Pem$ or $\mu^+\mu^-$)} &
{\Wprime $\to \cPq\cPaq\,\cPgn\cPagn $} & \\
  \hline
  700  & 1.267$\times 10^{-2}$ & 7.45$\times 10^{-2}$ & 1.35  \\
  800  & 6.815$\times 10^{-3}$ & 4.06$\times 10^{-2}$ & 1.35  \\
  900  & 3.842$\times 10^{-3}$ & 2.31$\times 10^{-2}$ & 1.34  \\
  1000 & 2.286$\times 10^{-3}$ & 1.39$\times 10^{-2}$ & 1.33 \\
  1100 & 1.413$\times 10^{-3}$ & 8.45$\times 10^{-3}$ & 1.32 \\
  1200 & 8.955$\times 10^{-4}$ & 5.34$\times 10^{-3}$ & 1.31  \\
  1300 & 5.750$\times 10^{-4}$ & 3.42$\times 10^{-3}$ & 1.30 \\
  1400 & 3.784$\times 10^{-4}$ & 2.27$\times 10^{-3}$ & 1.28  \\
  1500 & 2.554$\times 10^{-4}$ & 1.50$\times 10^{-3}$ & 1.26  \\
 \hline\hline
  \end{tabular}
  }
  \label{tab:sigSamples}
\end{table}

\begin{table}[htb]
  \topcaption{Background Monte Carlo samples. The
    notation $\ell$ stands for electrons, muons, or taus. The
    notation $\cPgn\cPagn$ includes all three neutrino flavors.}
  \centering{
  \begin{tabular}{|c|c|c|}
  \hline\hline
   & Cross Section  & \\
\raisebox{1.5ex}[0cm][0cm]{Channel}  & (pb)  &
\raisebox{1.5ex}[0cm][0cm]{Simulation Details} \\
  \hline
  \multicolumn{3}{|c|}{Dilepton Channels}  \\
\hline
  $\PW $+jets & 212.5  & LO ($\PT^\PW  > 100\GeV$)\\
  $t \bar t$ & 157.5  & NLO\\
  $\gamma V$+jets & 56.5 & LO\\
  $\cPZ/\gamma^*(\ell^+ \ell^-)$+jets & 25.1 & LO ($\PT^\cPZ > 100\GeV$) \\
  $\PW (\ell \nu)\,\PW (\ell \nu)$+jets & 3.8 & LO \\
  $\PW (\cPq \cPaq)\,\cPZ(\ell^+ \ell^-)$+jets & 1.14 & LO\\
  $\cPZ(\cPq\cPaq)\,\cPZ(\ell^+ \ell^-)$+jets & 0.57 & LO\\
\hline
  \multicolumn{3}{|c|}{\MET Channel}  \\
\hline
  QCD multijets & 5856.0  & LO ($500  < \HT < 1000\GeV$)\\
  QCD multijets & 122.6  & LO ($\HT > 1000\GeV$)\\
$Z/\gamma (\nu \bar \nu)$ + jets & 32.92 &  LO ($\HT > 200\GeV$)\\
  \hline\hline
  \end{tabular}
  }
  \label{tab:bkgSamples}
\end{table}

\section{Reconstruction and Event Selection} \label{reco}
Events are required to have at least one primary vertex of good
quality, where the vertex is reconstructed within $\pm$24\unit{cm} of the
nominal interaction point along
the beam axis, with a transverse distance from the beam spot of less
than 2\unit{cm} \cite{ref:PrimaryVertex}. The events are
reconstructed with the particle-flow (PF) technique
\cite{CMS-PAS-PFT-09-001}.
The PF algorithm reconstructs a complete
list of particle candidates in each event from the measurements in all
the components of the CMS detector in an integrated fashion. The
algorithm separately identifies muons, electrons, photons, charged and
neutral hadrons.
Charged hadrons that are consistent with
primary vertices other than the leading one (defined as the vertex with the
largest sum of track $\pt^2$)
are removed from the collection of particle
candidates used to reconstruct the jets, to mitigate the effects of
multiple proton-proton interactions within the same bunch crossing
(pileup).
Electrons are reconstructed as isolated objects in the
calorimeters which satisfy requirements on the shower
shape and the ratio of the hadronic to the electromagnetic
energy deposits.
Due to the boosted topology of this analysis, some care is needed when
reconstructing the $\cPZ \to \Pep\Pem$ decay: each
reconstructed electron
interferes with the isolation definition of the other electron and has to
be excluded from the isolation calculation in order to avoid
introducing inefficiencies.
The isolation criterion for electrons is the combined relative isolation $R_\text{iso}$
built upon information from the tracker, ECAL and HCAL. In calculating $R_\text{iso}$,
the track momenta and energy deposits,
excluding those associated with the electron itself,
are summed in a cone of radius $\Delta R < 0.3$ around the electron direction,
where $\Delta R \equiv \sqrt{(\Delta \phi)^2 + (\Delta \eta)^2}$,
and divided by the electron transverse momentum.
Muon tracks are built by combining a track from the inner tracker
and a track from the outer muon system.
No explicit isolation requirement is imposed on the muon candidates.
Lepton (electron and muon) candidates are required to
have a transverse (longitudinal) distance to the leading vertex smaller than 2
(5)\unit{mm}.
Jets are clustered from the reconstructed PF particles using the
infrared-safe anti-\kt~\cite{antikt} algorithm with
a distance
parameter of 0.7, as implemented in {\tt
FASTJET}~\cite{Cacciari:2005hq,Cacciari:2011ma}. The jet momentum is
determined as the vector sum of all particle momenta in the jet, and
is found in the simulated data to be within 5\% to 10\% of the true
momentum of the generator-level jet over the whole $\pt$ spectrum and
detector acceptance~\cite{JES2010}. An
area-based correction is applied to take into account the extra energy
clustered in jets due to additional proton-proton interactions within
the same bunch crossing, and for the average effect of out-of-time
pileup interactions~\cite{Fastjet1,Fastjet2}.
Jet energy corrections are derived from the energy balance of dijet and
photon+jet events.
Additional selection criteria are applied to each
event to remove spurious jet-like features originating from isolated
noise patterns from the hadron or the electromagnetic
calorimeters. The offline missing-transverse-momentum vector
($\vec{\pt}^{\text{miss}}$)
is calculated as the negative vector sum
of the transverse momenta of all
PF particles reconstructed in the event, and its magnitude is denoted by
$\MET$.

\subsection{Dilepton channels}
Candidate events are required to have at least two good quality
reconstructed leptons within the detector acceptance ($|\eta| <2.5$
for electrons and $|\eta| < 2.4$ for muons, with at least one muon
within $|\eta|<2.1$ at the trigger level) with $\PT > 45$\GeV.  We also
require at least one jet in the event reconstructed with $\PT > 30$\GeV within $|\eta|<2.4$.

Whenever two same-flavor, opposite-sign leptons are found in the event, a $\cPZ$
candidate is formed summing the four-momenta of the leptons.  We
select the $\cPZ$ candidates by requiring their invariant mass
to be in the range $70 < M_\cPZ < 110$\GeV
and with a transverse momentum
$\pt^\cPZ>150$\GeV.
If there are multiple $\cPZ$ candidates, the one with mass
closest to the nominal $\cPZ$ mass is selected. The requirement that the
dimuon mass is consistent with a $\cPZ \to \mu^{+} \mu^{-}$ decay
strongly suppresses non-prompt muons from jets.

The V candidate is selected by requiring a reconstructed jet
with $\PT >
250\GeV$ and $|\eta|<2.4$,
having an invariant mass $M_j$ (computed from the jet energy and momentum
calculated as the vector sum of the
four-momenta of the constituent PF particles) such that $65 < M_j <
120$\GeV.  We require the jet to be well separated from the two
leptons forming the $\cPZ$ candidate: $\Delta R \equiv \sqrt{(\Delta
\eta)^2 + (\Delta \phi)^2}>1.0$ for each lepton, where $\Delta \eta$ ($\Delta \phi$) is the
pseudorapidity (azimuthal) distance between the jet and the lepton
directions.
The selection has been optimized by maximizing the quantity $N_\mathrm{S}/\sqrt{N_\mathrm{S} + N_\mathrm{B}}$ (where $N_\mathrm{S}$ and $N_\mathrm{B}$ are the number of expected signal and
background events) for the lowest \Wprime mass point (700\GeV) considered
in this search.

Once the $\cPZ \to \ell \ell$ and (mono-jet) V$ \to \cPq\cPaq$ candidates
have been reconstructed, we combine their four momenta to compute the
mass of the parent particle, $M_{\mathrm{V}\cPZ}$. This variable is used to
evaluate the hypothesis of the signal presence in the
data sets analyzed.

\subsection{\texorpdfstring{\MET}{Missing ET} channel}

For the \MET channel, background from \PW-boson decays is reduced through rejection of
events with isolated electrons or muons with $\pt > 20\GeV$.
In order to further reduce leptonic backgrounds, we veto on the
presence of isolated tracks. For all tracks with $\pt > 10\GeV$ and $|\eta| < 2.4$, a
hollow cone of $0.02 < \Delta R < 0.30$ is constructed. The isolation
parameter of each track is defined as the scalar sum of all tracks with
$\pt > 1\GeV$ inside the cone, divided by the $\pt$ of the track.
Events containing a track with its isolation parameter smaller than 0.1 are discarded.
Events are then selected if the
jet with the highest transverse
momentum  has $\pt > 300\GeV$ and $|\eta| < 2.4$, and \MET is larger than
300\GeV.
In order to reduce the number of QCD multijet background events in the
signal region, events with more than two jets with $\pt > 30\GeV$ and
$|\eta| < 2.4$ are discarded. Events with exactly two jets above 30\GeV are
retained, if the azimuthal angle $\Delta\phi$ between the two jets is
smaller than 2.8\unit{radians}.
This condition improves the signal over background ratio by reducing the number of QCD dijet background events.

\begin{figure}[htb]
\includegraphics[width=0.5\textwidth]{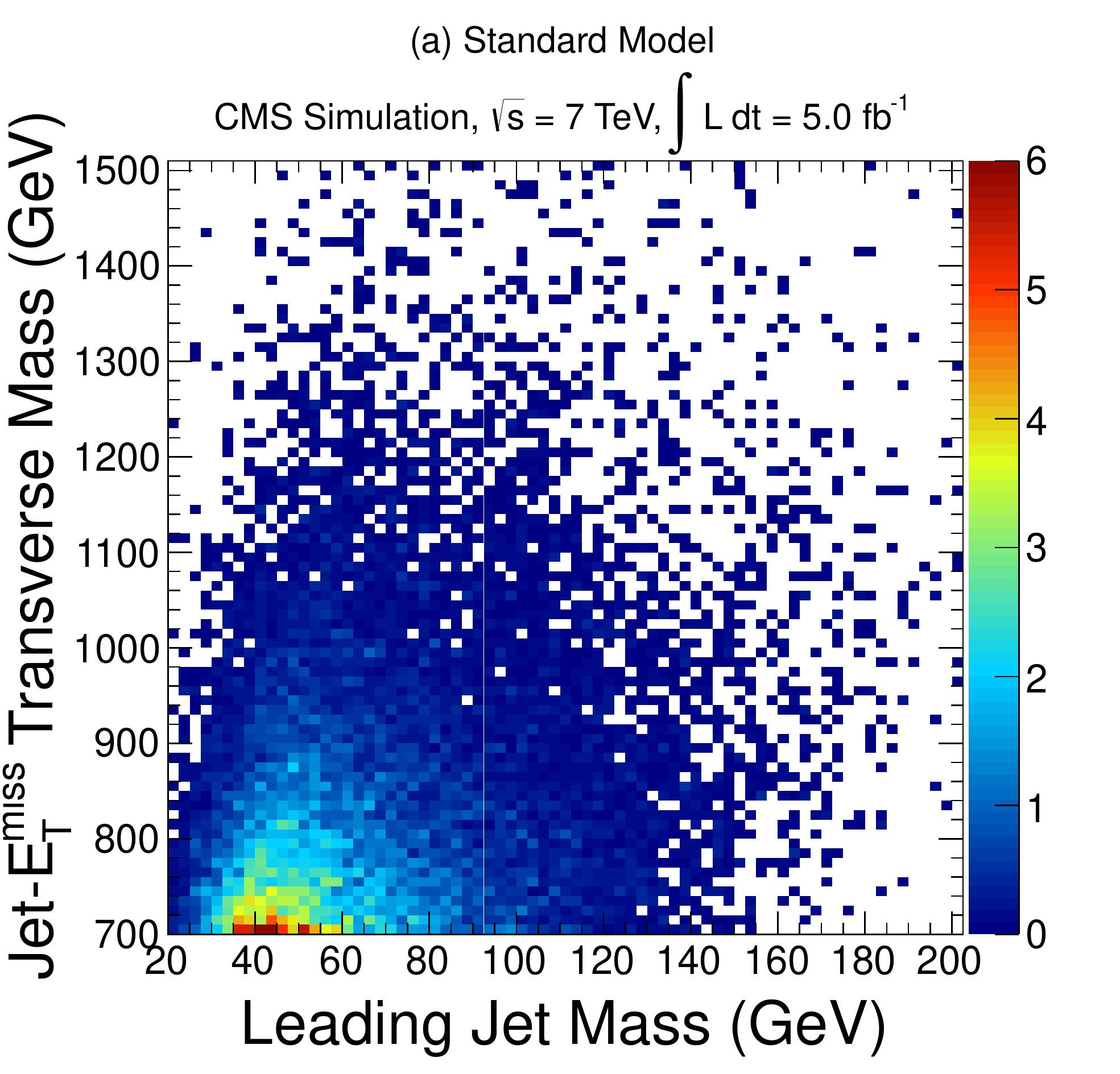}
\includegraphics[width=0.5\textwidth]{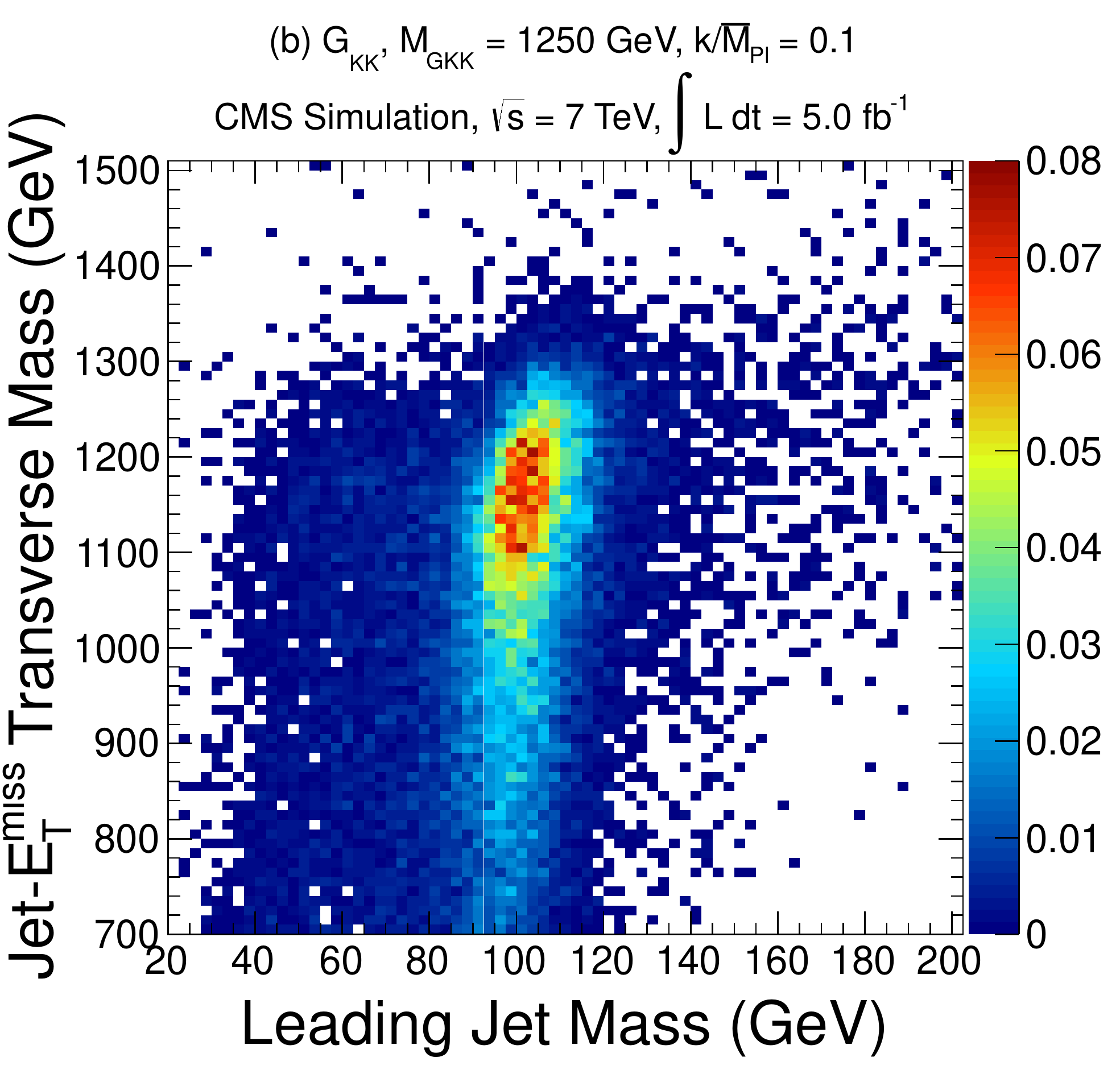}
\caption{Distributions of leading jet plus \MET
  transverse mass \vs leading jet mass for simulated standard model background sample (left)
  and RS graviton signal with
  $M_{\GKK} =  1250{\GeV}$ and $\kmpl = 0.05$ sample (right).
\label{fig:jetMassXGTM}}
\end{figure}

The signal sample is defined as the set of events that meet two extra requirements: the invariant mass of the leading jet, $M_j$, is larger than 70\GeV, and the jet-$\MET$ transverse mass, defined as
\[
	\gtm = \sqrt{ 2 \, \pt^{\text{jet}} \; \MET \; \Bigl[1 - \cos \Delta\phi(\text{jet},\pt^{\text{miss}} ) \Bigr] }  \;\;,
\]
is larger than 900\GeV.
Figure~\ref{fig:jetMassXGTM} on the left (right) shows the
two-dimensional $\gtm$ $vs$ $M_j$ distribution for the simulated SM
backgrounds (for a simulated signal sample with $M_{\GKK}$ = 1250\GeV).

In contrast to the approach used for the dilepton channels, here we perform a single ``event counting''
experiment by comparing the number of expected background and observed
events integrated over the region $M_j > 70$\GeV and $\gtm > 900$\GeV.

\section{Background Estimation}
We are discussing the background estimation separately for the dilepton and
\MET channels.
\subsection{Dilepton channels}
The analysis of the simulated data shows that the dominant (${\sim}90\%$) background after all selection requirements is the inclusive $\cPZ$
production (``$\cPZ$+jets''), with additional contributions from $\ttbar$+jets and the continuum SM diboson production ($\PW \cPZ$ and
$\cPZ\cPZ$). The shape and the overall normalization of the expected
background $M_{\mathrm{V}\cPZ}$ distributions are derived from data, with
additional cross-checks carried out with the inclusive simulated
background samples.  Effects caused by pileup are modeled by adding
to the generated events multiple proton-proton interactions with a multiplicity
distribution matched to the luminosity profile of the collision data.

The background is modeled using a control region consisting of a sideband in $M_j$ ($30 < M_j<65$\GeV).  The remaining selections are applied unmodified to these events, providing a sample that is kinematically equivalent to the nominal selection.
The robustness of this method against pileup effects, jet
  energy scale uncertainties, and variations in the sideband range has been
  confirmed with dedicated studies (Section~\ref{systematics}).

The procedure is as follows: we first produce the $M_{\mathrm{V}\cPZ}$ distribution for
the sideband
  selection. We define the ratio $\alpha(M_{\mathrm{V}\cPZ})$ as the
  total number of Monte Carlo background entries in the $M_{\mathrm{V}\cPZ}$ spectrum with
  the nominal (
  $65 < M_j<120$\GeV) and sideband (
  $30 < M_j<65$\GeV)  selections:

$$  \alpha(M_{\mathrm{V}\cPZ}) = \frac{N_\mathrm{NS}(M_{\mathrm{V}\cPZ})}{N_\mathrm{SB}(M_{\mathrm{V}\cPZ})} $$
where $N_\mathrm{NS}(M_{\mathrm{V}\cPZ})$ is the number of events in the
signal region and $N_\mathrm{SB}(M_{\mathrm{V}\cPZ})$ is the number of events in the
sideband region, contained in a bin of the V$\cPZ$ mass distribution
centered at a given value $M_{\mathrm{V}\cPZ}$.
We then use the product of the $M_{\mathrm{V}\cPZ}$ distribution made with the sideband selection in the data and the ratio $\alpha(M_{\mathrm{V}\cPZ})$ to derive an estimate of the background $M_{\mathrm{V}\cPZ}$ distribution with the nominal
selection.
Following the example of other resonance searches~\cite{Chatrchyan:2011ns}, we fit the $\alpha$-corrected sideband data $M_{\mathrm{V}\cPZ}$ distribution to
the following analytic function $f_A(M_{\mathrm{V}\cPZ})$, and the fit result is used to parametrize the expected SM background distribution:

$$f_A(M_{\mathrm{V}\cPZ}) = p_0
\frac{[1-(\frac{M_{\mathrm{V}\cPZ}}{\sqrt{s}})]^{p_1}}{\left(\frac{M_{\mathrm{V}\cPZ}}{\sqrt{s}}\right)^{\left[
    {p_2+p_3
    \log(\frac{M_{\mathrm{V}\cPZ}}{\sqrt{s}}) }\right] } }  \; ,$$
where $\sqrt{s}$ is the collision energy, $p_i$, $i=0,...,3$ are
free
parameters of the fit, and $M_{\mathrm{V}\cPZ}$ is expressed in\GeV. The fit determines both the
shape and the overall normalization of the expected background as a
function of $M_{\mathrm{V}\cPZ}$.
The fitting functions are then used to describe the expected background in
any subregion of the $M_{\mathrm{V}\cPZ}$ spectrum in the electron and muon channels.
There are several advantages in using the ratio $\alpha(M_{\mathrm{V}\cPZ})$ for the
background modeling of the $M_{\mathrm{V}\cPZ}$ distributions: the background
estimation becomes insensitive to effects such pile-up corrections and
integrated luminosity uncertainty which cancel out in the ratio;
$\alpha(M_{\mathrm{V}\cPZ})$ is less sensitive to improper modeling of the matrix
element calculation for the background and to theory systematics
(\eg normalization and factorization scale, PDFs, \etc) since the
background composition is similar in the two regions.

\begin{figure}[hbtp]
  \begin{center}
    \includegraphics[width=.49\textwidth]{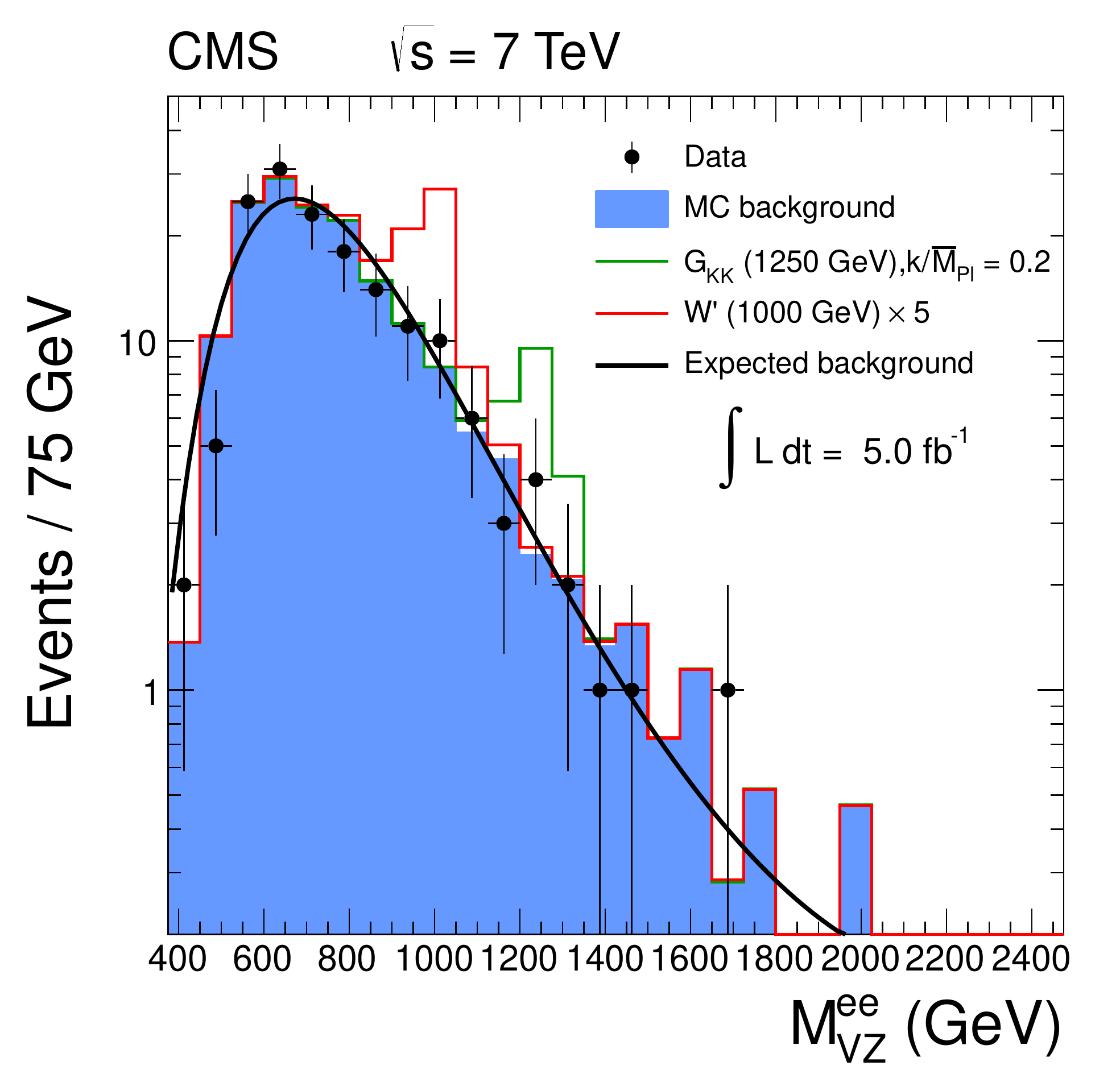}
    \includegraphics[width=.49\textwidth]{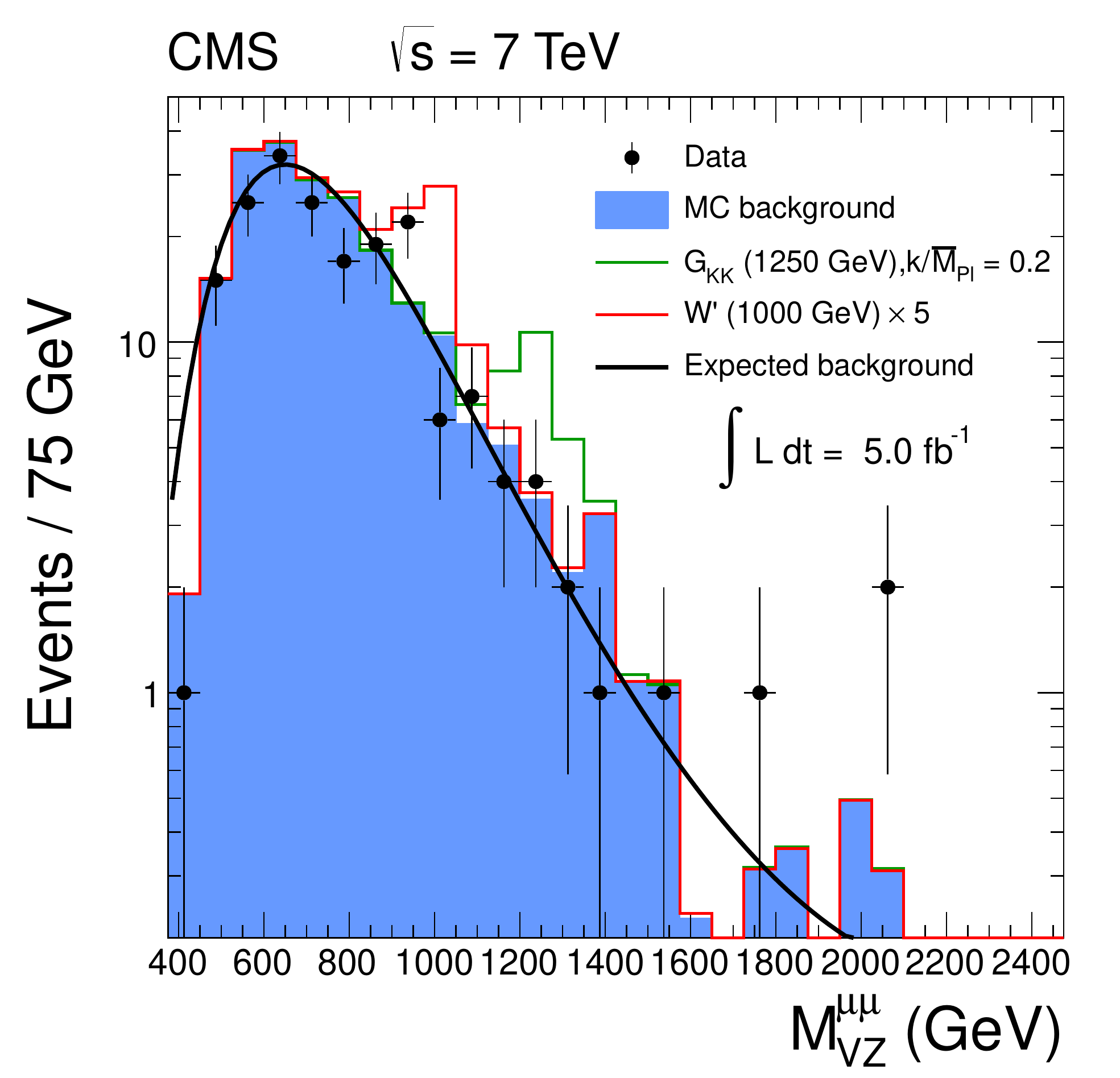}
    \caption{\label{fig:dataVsBgd} The comparison of the estimated
      background (black curve) with the
      total MC background (blue histogram) and the data (black points) for $M_{\mathrm{V}\cPZ}$  distributions for the electron (left) and the muon (right) channels.}
 \end{center}
\end{figure}

The comparison of the estimated background with the prediction from the
simulation and the data $M_{\mathrm{V}\cPZ}$ distributions is shown in Fig.~\ref{fig:dataVsBgd}.
No significant excess of events is observed, with the largest deviation appearing in the $\sim$900\GeV region in the muon channel.
The tail of the $M_{\mathrm{V}\cPZ}$ distribution, which is the
region of interest for the new resonance search, is well described by the fit.
A discrepancy is observed at low $M_{\mathrm{V}\cPZ}$ values. Any modeling imperfections, quantified as the difference between the best-fit function and the
MC simulation estimation,
are taken into account in the limit calculation by
assigning a systematic uncertainty.

\subsection{\texorpdfstring{\MET}{Missing ET} channel}
\label{sec:bgd_met}
By analyzing simulated data we determine that the dominant backgrounds in this channel after all
selections are inclusive $\cPZ \to \cPgn \cPagn$ ($\sim$70\%) and $\PW \to
\ell \cPgn$ ($\sim$30\%) production, with the charged lepton remaining undetected
 in the latter.
To estimate the SM background, we use a sideband-based technique similar to that described above, which utilizes events
that meet all other requirements but the $M_j$ and $\gtm$ thresholds.
In particular, the
events which meet all the selection requirements are classified into four
regions according to two
thresholds in jet mass and two thresholds in $\gtm$:
\begin{itemize}
\item Signal region A: $M_j > 70\GeV$, $\gtm >
  900\GeV$;
\item Sideband region B: $20 < M_j < 70\GeV$, $\gtm >
 900\GeV$;
\item Sideband region C: $20< M_j < 70\GeV$, $700 <
  \gtm  < 900\GeV$;
\item Sideband region D: $M_j >70\GeV$, $700 < \gtm  <
  900\GeV$;
  \end{itemize}

The numbers of events observed in the above regions are denoted as $N_\mathrm{A}$, ..., $N_\mathrm{D}$.

\begin{figure}[htb]
\includegraphics[width=0.5\textwidth]{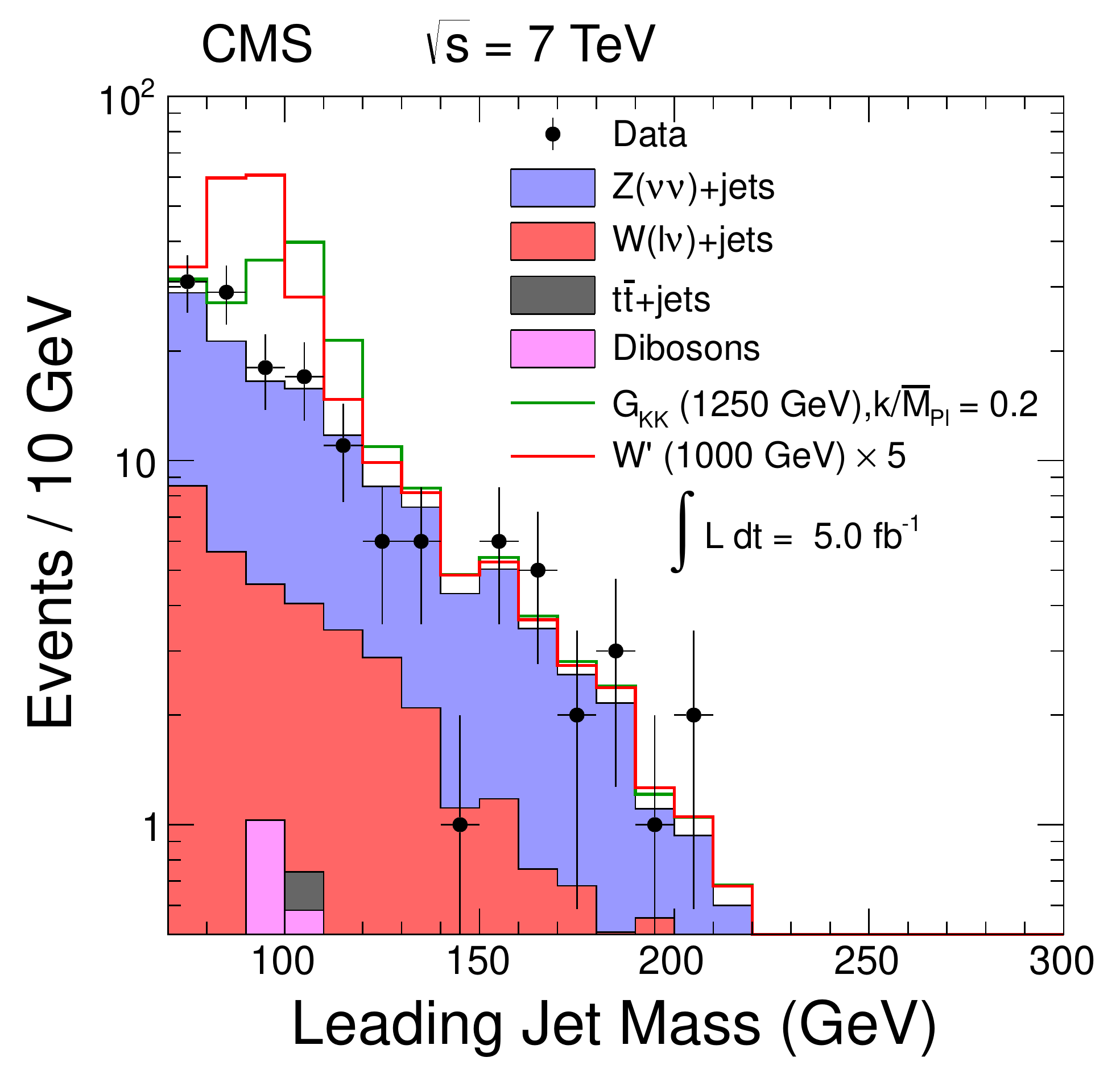}
\includegraphics[width=0.5\textwidth]{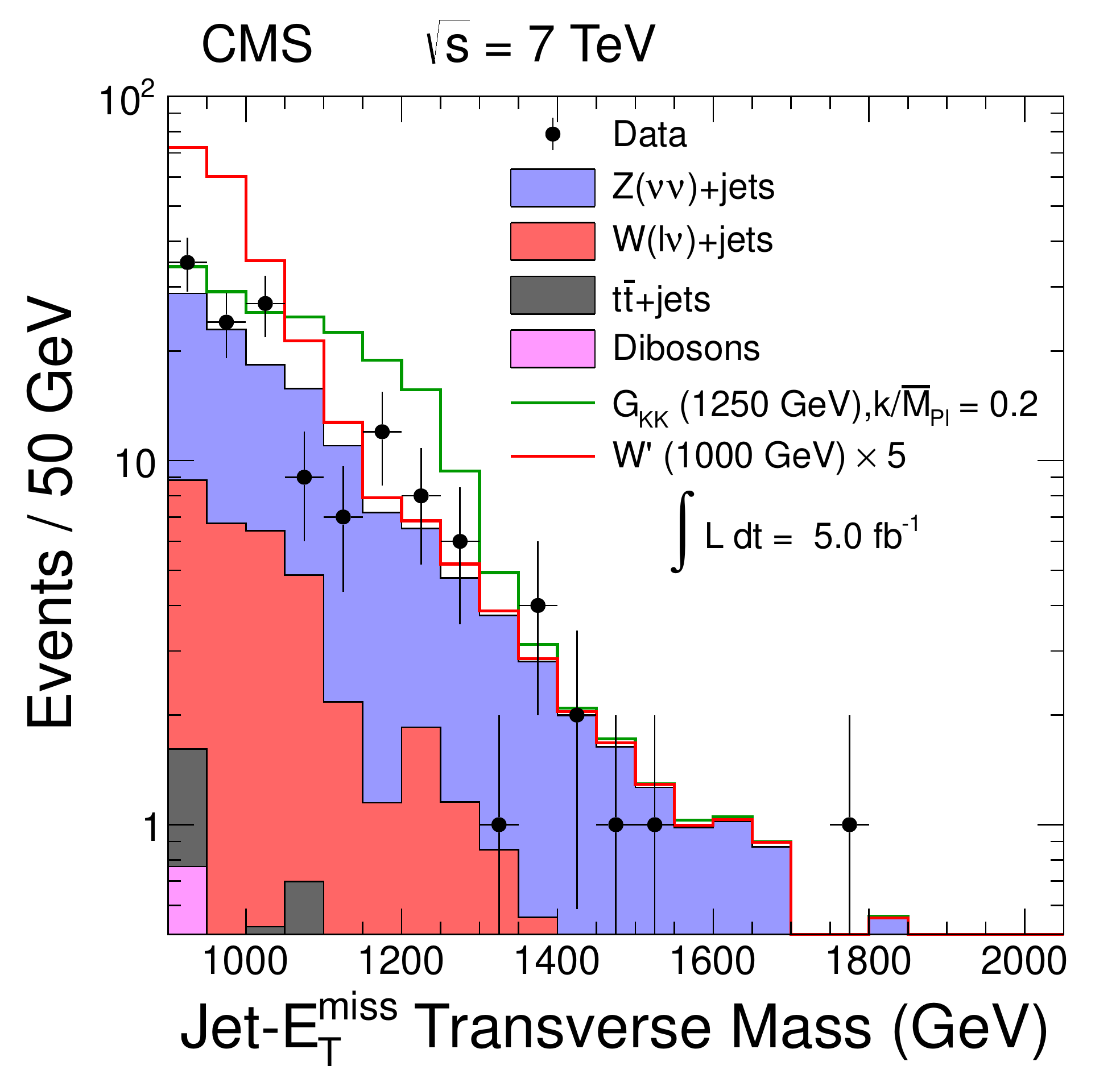}
\caption{Comparison between $\rho$-corrected simulated backgrounds
  and data in Region A for the leading jet mass (left) and jet-\MET{}
  transverse mass (right) distributions.}
\label{fig:dataAndBackgrounds_corrected}
\end{figure}

The estimated total background $B_{\text{est}}$ in Region A is given by the expression
\begin{equation}
\label{eq:estimatedBackgroundCorrected}
	B_{\text{est}} =  N_\mathrm{D} \cdot \frac{N_\mathrm{B}}{N_\mathrm{C}} \cdot \frac{1} {\rho}
\end{equation}
where $\rho$ is a correction factor to account for the correlation between
the jet mass and the jet-\MET transverse mass.
The $\rho$ parameter is estimated from the simulated SM samples by rearranging
Eq.~(\ref{eq:estimatedBackgroundCorrected}) in the following way:
\begin{equation}
\rho = \frac{N_\mathrm{D} \cdot N_\mathrm{B}}{N_\mathrm{A} \cdot N_\mathrm{C}}
\end{equation}
and setting the values of $N_\mathrm{A},\ldots,N_\mathrm{D}$
to the ones from the SM prediction.
Using the values reported in
Table~\ref{tab:yields_fourRegions} gives $\rho = 0.42 \pm 0.02$.
 The value of $\rho$ thus derived in then reinserted in Eq.~(\ref{eq:estimatedBackgroundCorrected}). Setting
$N_\mathrm{B},N_\mathrm{C},N_\mathrm{D}$ to the yields observed in the data, we obtain an estimate of the remaining background $B_{\text{est}} = 153 \pm 20$ events.
 Figures
\ref{fig:dataAndBackgrounds_corrected}~(left) and
\ref{fig:dataAndBackgrounds_corrected}~(right) show the comparison
between the simulated SM background in Region A (scaled to the estimated
value $B_{\text{est}}$---a scale factor of 11\%) and data, together with an example signal for the $M_j$
and $\gtm$ distributions.
There is agreement between the expected background and data
distributions.
\begin{table}[htb]
\centering
\topcaption{
Event yields for simulated SM samples, data, and the data/simulation ratio
in the four regions described in the text. The quoted uncertainties
include those due to the finite statistics of the simulated samples.
}
\label{tab:yields_fourRegions}
\begin{tabular} {| l | c | c | c | c |}
\hline\hline
Region & Yield & Data & SM Simulation & Data/Sim Ratio \\
\hline
\hline
A: signal & $N_\mathrm{A}$ 	& $138$		& $131 \pm 3$ & $1.05 \pm 0.02$ \\
B: sideband &$N_\mathrm{B}$  	& $125$	 	& $125 \pm 3$ & $1.00 \pm 0.03$\\
C: sideband & $N_\mathrm{C}$ 	& $542$		& $579 \pm 7$ & $0.94 \pm 0.01$\\
D: sideband & $N_\mathrm{D}$     & $283$		& $259 \pm 5$ & $1.09 \pm 0.02$\\
\hline\hline
\end{tabular}
\end{table}

\section{Systematic Uncertainties}\label{systematics}
The systematic uncertainties that are considered in this analysis can be divided
into two main categories: the uncertainty
in the determination of the SM background and the uncertainty in the
expected yields of signal events. All the systematic uncertainties are
summarized in Tables~\ref{tab:systematicZ}
and \ref{tab:systematicMET} for the dilepton and \MET channels,
respectively. The total
systematic uncertainty is the combination of the signal and background
systematic effects, assuming they
are completely uncorrelated.

\begin{table}[htbp]
\topcaption{Systematic uncertainties in the dilepton channels for given mass
  point and optimized mass window for the background
  (columns 3--6) and signal (columns 7--8) expected yields, following the
  procedures described in the text.
  In addition to the estimated signal
  uncertainties listed in the Table, constant uncertainties
  are considered on the integrated luminosity (2.2\%),
  the lepton reconstruction and trigger efficiencies as determined by the
  ``tag and probe'' method (2\%) \cite{ref:EWK-2010-002} and the V mass selection as determined from a sample
  of boosted $\ttbar$ events (9\%).  }
  \centering{
  \begin{tabular}{|c|c|c|c|c|c|c|c|}
  \hline\hline
\multicolumn{2}{|c|}{}  & \multicolumn{4}{c|}{Background}  &
\multicolumn{2}{c|}{Signal}
\\ \hline
Mass Point &   Mass Window & Stat. &  Fit & Diff. w/ & JES & JES & PDF\\
(\GeVns{})     &  (\GeVns{})       &  (\%)
& variations (\%) & MC (\%) & (\%)  & (\%) & (\%) \\
  \hline \hline
\multicolumn{8}{|c|}{\Wprime model, electron channel} \\ \hline
700 &  640-760   & 8 & 3 & 1 & 4 & 1 & 4 \\
800 &  755-845   & 8 & 21 & 1 & 5 & 1 & 5 \\
900 &  855-945   & 9 & 21 & 2 & 7 & 1 & 5 \\
1000 &  930-1070 & 11 & 17 & 5 & 2 & 1 & 6 \\
1100 &  1020-1180 & 12 & 22 & 4 & 3 & 1 & 6 \\
1200 &  1130-1270 & 15 & 26 & 6 & 7 & 1 & 6 \\
1300 &  1220-1380 & 17 & 46 & 6 & 41 & 1 & 7 \\
1400 &  1320-1480 & 20 & 64 & 28 & 14 & 1 & 8 \\
1500 &  1390-1610 & 23 & 72 & 31 & 26 & 1 & 9 \\
\hline \multicolumn{8}{|c|}{RS model, electron channel} \\ \hline
750 &  690-810   & 8 & 14 & 3 & 2 & 1 & 4 \\
1000 &  940-1060  & 11 & 17 & 5 & 7 & 1 & 6 \\
1250 &  1180-1320 & 16 & 36 & 3 & 8 & 1 & 7 \\
1500 &  1390-1610 & 23 & 72 & 3 & 26 & 1 & 9 \\
1750 &  1540-1960 & 31 & 64 & 10 & 48 & 1 & 10 \\
2000 &  1760-2240 & 42 & 42 & 26 & 110 & 1 & 11 \\
  \hline\hline
\multicolumn{8}{|c|}{\Wprime model, muon channel} \\ \hline
700 &  640-760   & 7 & 10 & 4 & 6 & 1 & 4 \\
800 &  755-845   & 8 & 19 & 6 & 8 & 1 & 5 \\
900 &  855-945   & 9 & 16 & 2 & 8 & 1 & 5 \\
1000 &  930-1070 & 10 & 15 & 8 & 1 & 1 & 6 \\
1100 &  1020-1180 & 12 & 20 & 8 & 1 & 1 & 6 \\
1200 &  1130-1270 & 14 & 29 & 21 & 2 & 1 & 7 \\
1300 &  1220-1380 & 16 & 43 & 30 & 6 & 1 & 7 \\
1400 &  1320-1480 & 19 & 50 & 68 & 10 &1 & 8 \\
1500 &  1390-1610 & 22 & 46 & 54 & 24 & 1 & 9 \\
\hline \multicolumn{8}{|c|}{RS model, muon channel} \\ \hline
750 &  690-810   & 7 & 14 & 2 & 7 & 1 & 4 \\
1000 &  940-1060  & 10 & 15 & 6 & 7 & 1 & 6 \\
1250 &  1180-1320 & 15 & 37 & 4 & 1 & 1 & 7 \\
1500 &  1390-1610 & 22 & 46 & 54 & 24 & 1 & 9 \\
1750 &  1540-1960 & 30 & 33 & 31 & 52 & 1 & 11 \\
2000 &  1760-2240 & 40 & 64 & 23 & 130 & 1 & 11 \\
  \hline\hline
  \end{tabular}
  }
  \label{tab:systematicZ}
\end{table}

\begin{table}[htb]
\centering
\topcaption{Systematic uncertainties in the \MET channel for the expected
  signal yields for the \Wprime mass range $M_{\Wprime} \in [700,1500]\GeV$ and graviton mass range $M_{\GKK} \in [750,2000]\GeV$.}
\label{tab:systematicMET}
\begin{tabular} {| c | c | c | c | c | }
\hline
\hline
Mass Point (\GeVns{}) & PDF (\%) & JES (\%) & \MET (\%) & Total (\%) \\
\hline
\hline \multicolumn{5}{|c|}{\Wprime model} \\ \hline
700 & 4 & 9 & 9 & 13 \\
800 & 5 & 8 & 8 & 12 \\
900 & 5 & 8 & 8 & 12 \\
1000 & 5 & 7 & 7 & 11 \\
1100 & 6 & 5 & 6 & 10 \\
1200 & 6 & 2 & 4 & 7 \\
1300 & 7 & 1 & 3 & 8 \\
1400 & 8 & 1 & 3 & 9 \\
1500 & 8 & 1 & 3 & 9 \\
\hline \hline
\multicolumn{5}{|c|}{RS model} \\ \hline
750 & 4 & 7 & 7 & 10 \\
1000 & 4 & 1 & 3 & 5 \\
1250 & 4 & 1 & 3 & 5 \\
1500 & 4 & 1 & 3 & 5 \\
1750 & 4 & 1 & 3 & 5 \\
2000 & 4 & 1 & 3 & 5 \\
\hline \hline
\end{tabular}
\end{table}

\subsection{Background systematic effects}
As we employ a method based on control samples in data for the background
determination, several
systematic effects are eliminated. In the following, we consider the remaining relevant
uncertainties in detail for the dilepton and \MET channels.

\subsubsection{Dilepton channels}
The expected number of background events in each
mass window is determined by the integral of the function $f_A(M_{\mathrm{V}\cPZ})$ in
the corresponding region. The statistical uncertainty is calculated by employing the full
covariance error matrix of the fit parameter uncertainties in the integral of the
fitting function in the mass window.
The pileup and jet
energy scale (JES) systematics can potentially affect the background
determination through the $\alpha(M_{\mathrm{V}\cPZ})$ ratio and are considered
separately. The former is found to have a negligible effect. For the
latter, the uncertainty is
evaluated by
varying the jet \PT to $\PT \pm \sigma_\mathrm{JES}(\PT,\eta)$, where $\sigma_\mathrm{JES}(\PT,\eta)$ is the total jet uncertainty, and applying the full fitting procedure.
The yield differences, in each mass window,
between the expected background
with the positive
($N_{+\text{Bgd}}$) and negative
($N_{-\text{Bgd}}$) jet energy scale variation
with respect to the nominal selection and fit
are taken as the ${\pm}1\,\sigma$ estimates for the JES systematic
uncertainty.
We also consider several variations in
the fitting procedure (fitting range, functional form, and sideband
definition).
These variations are compared to the difference in the number of expected
background events in the mass window as estimated from data and with MC
simulation. The largest of the two is used as
the systematic uncertainty in the background determination.

\subsubsection{\MET channel}
\label{sec:syst_MET_channel}

To evaluate the robustness of the evaluation of the expected background
two tests are conducted.

The first test studies the dependence of the
correction factor $\rho$ on the definition of the sideband regions.
We vary the definition of the sideband regions by changing the thresholds in the $\jetmass$ and
$\gtm$ variables in the intervals 20--70\GeV and 650--750\GeV, respectively.
We find that the resultant variation in the mean estimated background is typically
5\% or less, confirming the robustness of the sideband method.

A second test is used to check the propagation of all the uncertainties involved in the
$B_{\text{est}}$ calculations.
We generate a series of pseudo-experiments with the number of events constrained to be equal to
that of the actual experiment.  We obtain a value of $\rho$ and calculate the mean estimated background
in each case.  The distribution of the values of $B_{\text{est}}$ thus obtained has a variance of 20
events.  This result is in agreement with the estimated uncertainty on $B_{\text{est}}$ obtained in
Section 5.2, using the yields of $N_\mathrm{B}$, $N_\mathrm{C}$ and $N_\mathrm{D}$ observed in the data.

The mean expected SM background in Region A, within the uncertainties
calculated above, is compatible with the observed event yield in the signal
region, $N_\mathrm{A} = 138$ events.

\subsection{Signal systematics}
There are several systematic uncertainties in the expected signal yields that
are common across channels. These uncertainties are
on the luminosity
measurement, the JES effects on jets, the
PDF, and the trigger and reconstruction
efficiencies.
A value of 2.2\% was taken for the uncertainty in the measurement of the integrated luminosity~\cite{LUMIPAS}.

To determine the effect of JES uncertainty,
we vary the jet \PT to $\PT \pm \sigma_\mathrm{JES}(\PT,\eta)$, where $\sigma_\mathrm{JES}(\PT,\eta)$ is the total jet uncertainty, and apply the full analysis selection.
The differences in the signal yields
$N_{+\text{sig}}$ and $N_{-\text{sig}}$ with respect to the nominal
selection $N_\text{sig}$ are taken as the ${\pm}1\,\sigma$
estimates for the JES systematic uncertainty.
For \Wprime and RS signals
with the mass in the  range [700, 2000]\GeV in the dilepton channels
and in the range [1250, 2000]\GeV in the \MET channel, this systematic uncertainty is
less than 1\%.
However, for resonance masses in the range [700, 1200]\GeV in the \MET
channel this systematic uncertainty is found to be between 2 and 9\%, owing to threshold
effects.
To estimate the systematic uncertainty
associated with the choice of the PDF used for the simulated samples,
  three scenarios are considered:
{CTEQ6.6}, {MSTW2008} and {NNPDF2.0}~\cite{LHAPDF}. The systematic
uncertainty is set to half of the difference between the maximum and the
minimum PDF values predicted for each mass point~\cite{PDF4LHC}.

\subsubsection{Dilepton channels}
To account for differences in trigger and reconstruction efficiencies
between the Monte Carlo simulation and data, we determine scaling factors by
using data control samples of $\cPZ \to \mu\mu$ and $\cPZ \to \Pe\Pe$
candidate events~\cite{CMS_MUO_2010_004, EWK-10-002-PAS}. We derive
corrections for the muon ($0.974 \pm
0.001$) and the electron ($0.960 \pm 0.004$) channels and we apply them to
the expected signal yields.
These
numbers assume
that the efficiency does not vary with \pt (\ET). However,
we observe
a small decrease (increase) in the efficiency in the asymptotic high-\pt (high-\ET) region for muons
(electrons) of about 2\%.
This small difference is used as the
systematic uncertainty in the expected number of signal events for each
mass point considered in this study. Finally, we assign a 9\% systematic uncertainty on the V mass selection
efficiency. This is determined by studying an independent sample of boosted $\ttbar
\rightarrow \PW  \cPqb\, \PW  \cPqb$ events in which one of the $\PW $ bosons decays
leptonically and the other hadronically.

\subsubsection{\MET channel}
Propagating the jet energy scale effects to the calculation of \MET,
and accounting for the anticorrelation between jets and  \MET itself,
we estimate a systematic effect of around 3\% for all values of
$\gravitonmass$ studied, except for the lowest $\gravitonmass =
750\GeV$. In this case, because of threshold effects, the
systematic effect is found to be around 7\%.

Summing in quadrature the uncertainties above, we arrive at a final 5\%
systematic uncertainty on the signal acceptance and efficiency except for
$\gravitonmass = 750\GeV$, where a value of
10\% is obtained for the final systematic uncertainty on the signal acceptance and efficiency.

\section{Results}
We do not observe any significant excess over the expected background.
We employ the modified frequentist CL$_\mathrm{S}$ statistical
method \cite{LEP-CLs,Junk:hep-ex9902006} to
search for exotic V$\cPZ$ resonances. For the dilepton channels we use a
series of search windows corresponding to different mass hypotheses.
Each mass window is optimized to give the best exclusion limit, a procedure
which is also appropriate for establishing a new resonance discovery.
The mass windows optimization has been carried out separately for the \Wprime
and RS graviton hypotheses to account for differences in the width and
efficiencies.
For the \MET channel we perform a single counting experiment in the $\gtm >
900\GeV$ and $M_j > 70\GeV$ region.
We calculate $95\%$ confidence level (CL) exclusion
limits on the combined products of the cross section times the branching ratio
$\sigma (\Pp\Pp \to \Wprime) \times \mathcal{B} (\Wprime \to \PW \cPZ)$ and
$\sigma (\Pp\Pp \to \GKK) \times \mathcal{B} (\GKK \to \cPZ \cPZ)$
for the three final states under study (separately
and combined) as a function of the mass of the
hypothetical resonance. We interpret these exclusion limits in two benchmark signal models:
SSM \Wprime and RS graviton.

\begin{figure}[hbtp]
  \begin{center}
    \includegraphics[width=.75\textwidth]{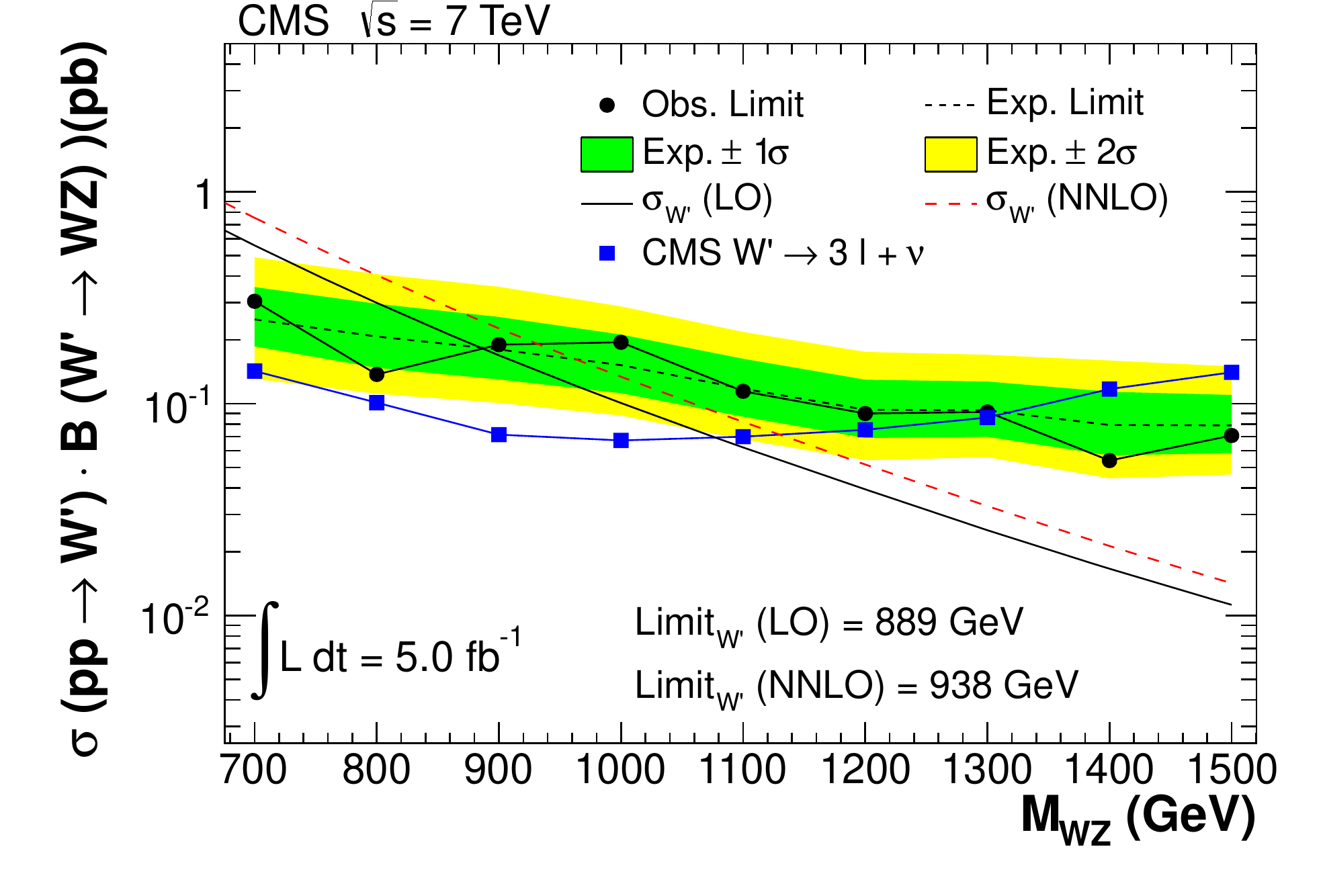}
    \includegraphics[width=.75\textwidth]{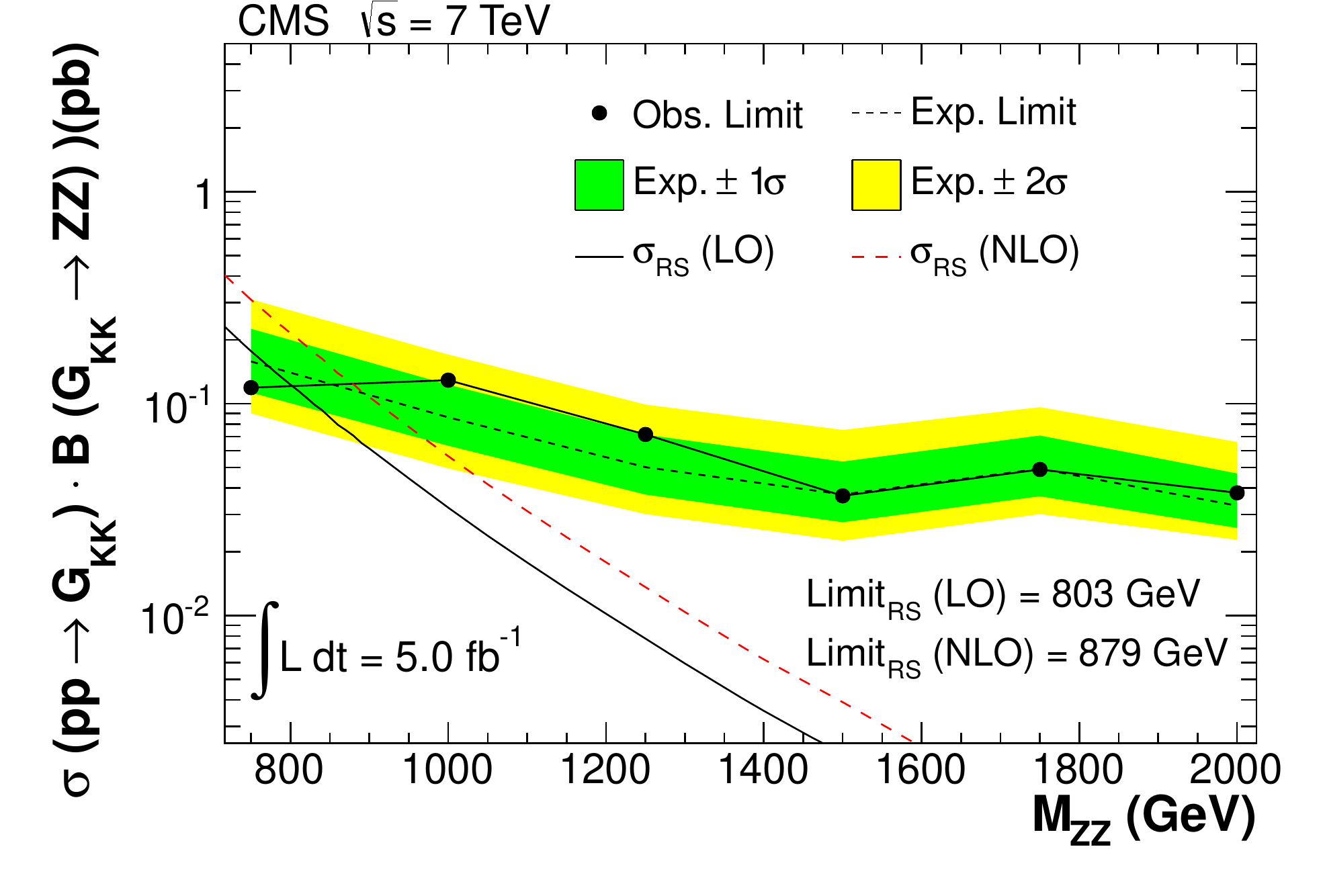}
    \caption{\label{fig:limitCombinedData} Observed and expected 95\% CL upper
     cross section limits
      and comparison with the theoretical predictions
      in \Wprime (top) and RS
      graviton with $\kmpl = 0.05$ (bottom) models for the combination of electron, muon,
      and \MET channels.
      The limits are calculated with the modified frequentist CL$_{\text{S}}$ statistical
      method.}
 \end{center}
\end{figure}

\begin{figure}[hbt]
  \begin{center}
    \includegraphics[width=.75\textwidth]{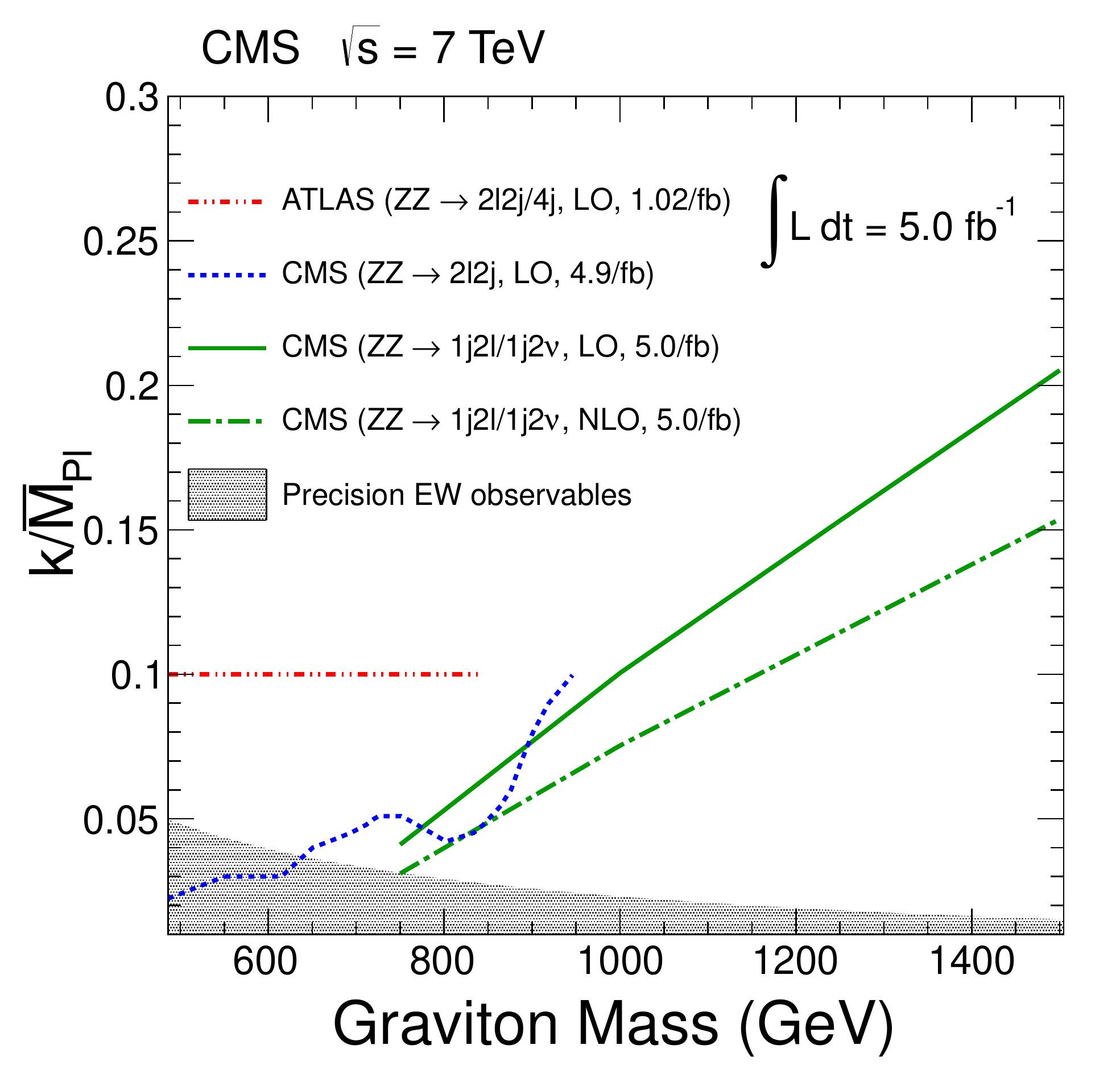}
    \caption{\label{fig:RS2DPlot} Observed exclusion limits from this
      analysis (ZZ $\rightarrow 1j\,2\ell/1j\,2\nu$) interpreted in the context of
    the RS graviton model, assuming a LO prediction (solid green line) and NLO prediction (dot-dashed green line).
    Also shown are results from
    ATLAS (ZZ $\rightarrow 2\ell\,2j/4j$) \cite{Collaboration:2012iua} (dot-dot-dashed red line) and
    another CMS publication (ZZ $\rightarrow 2\ell\,2j$) \cite{CMS-PAS-EXO-11-102} (dotted blue line).
    The shaded region corresponds to the indirect limits derived from
    precision electroweak observables in \cite{Davoudiasl:2000wi}.
}
 \end{center}
\end{figure}

The limit setting is performed by looking for an excess over the expected
background in the V$\cPZ$ mass distributions for the three channels separately.
Tables~\ref{tab:eYields} and \ref{tab:mYields} show the search windows
 for each mass point with the corresponding signal efficiency and the
 numbers of expected background and observed events in the electron and muon
 channels, respectively. These numbers are used as input for the
 calculation of  the expected and observed exclusion limits on cross section times branching ratios
 at 95$\%$ CL that are also reported in the same Tables.
Table~\ref{tab:metYields} shows the signal efficiency and the observed
and expected exclusion limits as a function of the signal mass
in the \MET  channel. The combined results are reported in
 Table~\ref{tab:combYields}.
The exclusion limits as a function of the V$\cPZ$ resonance mass can be seen
in Fig.~\ref{fig:limitCombinedData}, where a linear interpolation is used
between the benchmark mass values.
These limits can be interpreted in the
theoretical framework of the \Wprime and RS graviton models.
We exclude SSM \Wprime
bosons with masses
in the range 700--940 (890)\GeV in the SSM at NNLO (LO)
at 95\% CL
These results are complementary to the ones obtained in the tri-lepton
analysis (with $M_{\Wprime} > 1143$\GeV in the SSM \cite{CMS_EXO_11_041}).
The exclusion limit calculated in the RS
 graviton model rules out masses ($M_{\GKK}$)
 in the range 750--880 (800)\GeV
 for \kmpl = 0.05 at NLO (LO). Assuming the resonance width is much smaller than the experimental resolution for the range of
\kmpl considered here, the limit can be translated into the $M_{\GKK}$-\kmpl\,
plane. We do this by using the quadratic dependence of
the cross section on \kmpl, and by assuming that the signal
efficiency remains the same.
The result is shown in Fig.~\ref{fig:RS2DPlot}.

These results are particularly relevant in the context of
RS models proposed in recent studies \cite{Agashe:2007zd},
with SM fields  propagating in the extra dimension
where the graviton
coupling to light fermions is strongly suppressed.
This opens the possibility to an enhancement of
the branching fractions for final states with V pairs,
and motivates the investigation of large values
of \kmpl \cite{Agashe:2007zd}.
In this scenario the previously-published searches for RS gravitons decaying to
$\gamma \gamma$ and $\ell^+ \ell^-$ final states~\cite{Chatrchyan:2011fq}
do not impose stringent bounds, since the branching fraction for these final states is suppressed.
The results derived in this analysis are currently the most stringent in the V pair channel, and provide important constraints
that are complementary to the ones from the search for resonances decaying to boosted top pairs~\cite{boostedTop}.
\begin{table}[htbp]
\topcaption{Electron channel: Search window for each mass point with the
  corresponding signal efficiency (``$\epsilon_\text{sig}$'') and the
  numbers of mean expected background
  (``$N_\text{bgd}$'') and observed (``$N_\text{obs}$'') events. The
  uncertainties include both statistical and systematic effects. These numbers are used
  as input for the calculation of  the expected and observed exclusion
  limits on
  $\sigma (\Pp\Pp \to \Wprime) \times \mathcal{B} (\Wprime \to \PW \cPZ)$ and
 $\sigma (\Pp\Pp \to \GKK) \times \mathcal{B} (\GKK \to \cPZ \cPZ)$
  at 95$\%$ CL which are reported in the
  last two columns.}
\centering{
\begin{tabular}{|c|c|c|c|c|c|c|}
\hline\hline
Mass Point & Window &  &  & $\epsilon_\text{sig}$
 & Obs. Limit & Exp. Limit \\
(\GeV) & (\GeV) & \raisebox{1.5ex}[0cm][0cm]{$N_\text{bgd}$} &
\raisebox{1.5ex}[0cm][0cm]{$N_\text{obs}$} & ($\%$) & (pb) & (pb)
\\ \hline \hline
\multicolumn{7}{|c|}{\Wprime model} \\ \hline
700 & 640-760 & 39.7 $\pm$ 3.9 & 43       & 37$\pm$4 & 0.44  &   0.37   \\
800 & 755-845 & 24.6 $\pm$ 5.7 & 23       & 36$\pm$4 & 0.33 &   0.35   \\
900 & 855-945 & 17.1 $\pm$ 4.2 & 12       & 40$\pm$4 & 0.18  &   0.24   \\
1000 & 930-1070 & 17.1 $\pm$ 3.5 & 17   & 49$\pm$5 & 0.20  &   0.20 \\
1100 & 1020-1180 & 12.0 $\pm$ 3.0 & 13 & 48$\pm$5 & 0.20  &   0.18 \\
1200 & 1130-1270 & 6.3 $\pm$ 1.9 & 5     & 41$\pm$5 & 0.13 &   0.15 \\
1300 & 1220-1380 & 4.4 $\pm$ 2.8 & 6     & 32$\pm$4 & 0.25 &   0.20 \\
1400 & 1320-1480 & 2.7 $\pm$ 1.8 & 2     & 23$\pm$3 & 0.18 &   0.19  \\
1500 & 1390-1610 & 2.5 $\pm$ 2.0 & 2     & 19$\pm$2 & 0.22 &   0.22  \\ \hline
\multicolumn{7}{|c|}{RS model} \\ \hline
750 & 690-810 & 37.1 $\pm$ 6.0 & 32     & 27$\pm$3 & 0.21  &   0.25   \\
1000 & 940-1060 & 14.6 $\pm$ 3.1 & 16 & 35$\pm$4 & 0.14  &   0.13 \\
1250 & 1180-1320 & 4.9 $\pm$ 1.9 & 7    & 35$\pm$4 & 0.11 &   0.08   \\
1500 & 1390-1610 & 2.5 $\pm$ 2.0 & 2    & 27$\pm$3 & 0.08 &   0.08   \\
1750 & 1540-1960 & 2.0 $\pm$ 1.7 & 1    & 16$\pm$2 & 0.10 &   0.12   \\
2000 & 1760-2240 & 1.3 $\pm$ 1.6 & 0    & 17$\pm$2 & 0.06 &   0.05   \\
\hline \hline
\end{tabular}
}
\label{tab:eYields}
\end{table}

\begin{table}[htbp]
\topcaption{Muon channel: Search window for each mass point with the
  corresponding signal efficiency (``$\epsilon_\text{sig}$'') and the
  numbers of mean expected background
  (``$N_\text{bgd}$'') and observed (``$N_\text{obs}$'') events. The
  uncertainties include both statistical and systematic effects. These
  numbers are used as
  input for the calculation of the expected and observed exclusion limits
  on
  $\sigma (pp \to \Wprime) \times \mathcal{B} (\Wprime \to \PW \cPZ)$ and
  $\sigma (pp \to \GKK) \times \mathcal{B} (\GKK \to \cPZ \cPZ)$
   at 95$\%$ CL, which
  are reported in the last two columns.}
\centering{
\begin{tabular}{|c|c|c|c|c|c|c|}
\hline
Mass Point & Window &  &  & $\epsilon_\text{sig}$
 & Obs. Limit & Exp. Limit \\
(\GeV) & (\GeV) & \raisebox{1.5ex}[0cm][0cm]{$N_\text{bgd}$} &
\raisebox{1.5ex}[0cm][0cm]{$N_\text{obs}$} & ($\%$) & (pb) & (pb)
\\ \hline \hline
\multicolumn{7}{|c|}{\Wprime model} \\ \hline
700 & 640-760 & 48.7 $\pm$ 8.9 & 45 & 40$\pm$4 & 0.40  &   0.45   \\
800 & 755-845 & 28.6 $\pm$ 6.9 & 21 & 40$\pm$4 & 0.25 &   0.34   \\
900 & 855-945 & 19.2 $\pm$ 4.3 & 23 & 41$\pm$4 & 0.37  &   0.29   \\
1000 & 930-1070 & 18.7 $\pm$ 3.7 & 26 & 51$\pm$6 & 0.34  &   0.22 \\
1100 & 1020-1180 & 12.9 $\pm$ 3.1 & 12 & 52$\pm$6 & 0.17  &   0.18 \\
1200 & 1130-1270 & 6.7 $\pm$ 2.2 & 8 & 44$\pm$5 & 0.18 &   0.15 \\
1300 & 1220-1380 & 4.6 $\pm$ 2.1 & 4 & 42$\pm$5 & 0.13 &   0.13 \\
1400 & 1320-1480 & 2.9 $\pm$ 2.0 & 1 & 39$\pm$5 & 0.08 &   0.11  \\
1500 & 1390-1610 & 2.6 $\pm$ 1.7 & 2 & 40$\pm$5 & 0.11 &   0.11  \\ \hline
\multicolumn{7}{|c|}{RS model} \\ \hline
750 & 690-810 & 44.1 $\pm$ 9.2 & 34 & 30$\pm$3 & 0.19  &   0.26   \\
1000 & 940-1060 & 15.9 $\pm$ 3.4 & 20 & 39$\pm$4 & 0.17  &   0.13 \\
1250 & 1180-1320 & 5.2 $\pm$ 2.1 & 6 & 41$\pm$5 & 0.08 &   0.07   \\
1500 & 1390-1610 & 2.6 $\pm$ 1.7 & 2 & 44$\pm$6 & 0.05 &   0.05   \\
1750 & 1540-1960 & 2.1 $\pm$ 1.4 & 2 & 32$\pm$5 & 0.06 &   0.06   \\
2000 & 1760-2240 & 1.3 $\pm$ 1.9 & 2 & 42$\pm$6 & 0.06 &   0.05   \\ \hline\hline
\end{tabular}
}
\label{tab:mYields}
\end{table}

\begin{table}[htbp]
\topcaption{\MET channel: Expected and observed exclusion limits
  on
  $\sigma (\Pp\Pp \to \Wprime) \times \mathcal{B} (\Wprime \to \PW \cPZ)$ and
  $\sigma (\Pp\Pp \to \GKK) \times \mathcal{B} (\GKK \to \cPZ \cPZ)$
  at 95$\%$
  CL for each mass point with the corresponding signal efficiency (``$\epsilon_\text{sig}$''). In the
  $\gtm > 900\GeV$ region the
  expected background is $B_{\text{est}} = 153 \pm 29$, including both
  statistical and systematic uncertainties, and the number of observed
  events is 138. These parameters are common for all mass points considered
  in this channel.}
\centering{
\begin{tabular}{|c|c|c|c|}
\hline
Mass Point (\GeVns{}) & $\epsilon_\text{sig}$ ($\%$) & Obs. Limit (pb) & Exp. Limit (pb) \\ \hline \hline
\multicolumn{4}{|c|}{\Wprime model} \\ \hline
700  & 0.2$\pm$0.1 & 29    & 33 \\
800  & 0.9$\pm$0.1 & 7.0   & 8.2 \\
900  & 8.0$\pm$0.5 & 0.77 & 0.90 \\
1000 & 31$\pm$2 & 0.19 & 0.23 \\
1100 & 49$\pm$2 & 0.13 & 0.15 \\
1200 & 58$\pm$3 & 0.11 & 0.13 \\
1300 & 64$\pm$3 & 0.10 & 0.11 \\
1400 & 66$\pm$3 & 0.09 & 0.11 \\
1500 & 69$\pm$3 & 0.09 & 0.11 \\ \hline
\multicolumn{4}{|c|}{RS model} \\ \hline
750  & 0.7$\pm$0.1 & 4.1  &  4.8 \\
1000 & 25$\pm$2 & 0.12 & 0.14 \\
1250 & 43$\pm$3 & 0.07 & 0.08 \\
1500 & 54$\pm$3 & 0.06 & 0.07 \\
1750 & 60$\pm$3 & 0.05 & 0.06 \\
2000 & 63$\pm$3 & 0.05 & 0.06 \\ \hline
\end{tabular}
}
\label{tab:metYields}
\end{table}

\begin{table}[htbp]
\topcaption{\label{tab:combYields}Combined channels: Expected and observed
  exclusion
  limits on
   $\sigma (\Pp\Pp \to \Wprime) \times \mathcal{B} (\Wprime \to \PW \cPZ)$ and
  $\sigma (\Pp\Pp \to \GKK) \times \mathcal{B} (\GKK \to \cPZ \cPZ)$
  at 95$\%$ CL for the electron, muon, and \MET channels combined for each mass
  point and search window.}
\centering{
\begin{tabular}{|c|c|c|c|}
\hline
Mass Point (\GeVns{}) & Window & Obs. Limit (pb) & Exp. Limit (pb) \\ \hline \hline
\multicolumn{4}{|c|}{\Wprime model} \\ \hline
700 & 640--760 &  0.30  &   0.25    \\
800 & 755--845 &  0.14 &    0.21   \\
900 & 855--945 &  0.19  &   0.18   \\
1000 & 930--1070 & 0.19  &   0.15 \\
1100 & 1020--1180 &  0.11 &   0.12 \\
1200 & 1130--1270 &  0.09 &   0.09 \\
1300 & 1220--1380 &  0.09 &   0.09 \\
1400 & 1320--1480 &  0.05 &   0.08  \\
1500 & 1390--1610 &  0.07 &   0.08  \\ \hline
\multicolumn{4}{|c|}{RS model} \\ \hline
750 & 690--810 &  0.12  &   0.16   \\
1000 & 940--1060 &  0.13  &   0.09 \\
1250 & 1180--1320 &  0.07 &   0.05   \\
1500 & 1390--1610 &  0.04 &   0.04   \\
1750 & 1540--1960 &  0.05 &   0.05   \\
2000 & 1760--2240 &  0.04 &   0.03   \\
\hline \hline
\end{tabular}
}
\end{table}

\section{Summary}
A search for new exotic particles decaying to the V$\cPZ$ final state was performed,
where V is either a $\PW$ or a $\cPZ$ decaying to hadrons, and the $\cPZ$ decays
to electrons, muons, or a neutrino pair.
The analysis is based on a data sample of pp collisions corresponding to an integrated luminosity of \usedLumiVZ collected by the CMS experiment at the LHC at $\sqrt{s}=7$\TeV in 2011.
No significant excess is observed in the mass distribution of the V$\cPZ$ candidates compared with the background
expectation from standard model processes. Lower bounds at the $95\%$ confidence level are set in two theoretical models on the mass
of hypothetical particles decaying to the V$\cPZ$ final state.
Assuming heavy charged vector bosons in the sequential standard model, \Wprime bosons are excluded with masses
in the range 700--940 (890)\GeV
at NNLO (LO). In the Randall--Sundrum model, graviton resonances with masses
in the range 750--880 (800)\GeV
at NLO (LO) are excluded for \kmpl=0.05.
These are the first results from the LHC on V$\cPZ$ searches using final states with a
boosted massive jet and a lepton pair or missing transverse energy.

\section*{Acknowledgements}
We congratulate our colleagues in the CERN accelerator departments for the excellent performance of the LHC and thank the technical and administrative staffs at CERN and at other CMS institutes for their contributions to the success of the CMS effort. In addition, we gratefully acknowledge the computing centres and personnel of the Worldwide LHC Computing Grid for delivering so effectively the computing infrastructure essential to our analyses. Finally, we acknowledge the enduring support for the construction and operation of the LHC and the CMS detector provided by the following funding agencies: BMWF and FWF (Austria); FNRS and FWO (Belgium); CNPq, CAPES, FAPERJ, and FAPESP (Brazil); MEYS (Bulgaria); CERN; CAS, MoST, and NSFC (China); COLCIENCIAS (Colombia); MSES (Croatia); RPF (Cyprus); MoER, SF0690030s09 and ERDF (Estonia); Academy of Finland, MEC, and HIP (Finland); CEA and CNRS/IN2P3 (France); BMBF, DFG, and HGF (Germany); GSRT (Greece); OTKA and NKTH (Hungary); DAE and DST (India); IPM (Iran); SFI (Ireland); INFN (Italy); NRF and WCU (Korea); LAS (Lithuania); CINVESTAV, CONACYT, SEP, and UASLP-FAI (Mexico); MSI (New Zealand); PAEC (Pakistan); MSHE and NSC (Poland); FCT (Portugal); JINR (Armenia, Belarus, Georgia, Ukraine, Uzbekistan); MON, RosAtom, RAS and RFBR (Russia); MSTD (Serbia); SEIDI and CPAN (Spain); Swiss Funding Agencies (Switzerland); NSC (Taipei); ThEP, IPST and NECTEC (Thailand); TUBITAK and TAEK (Turkey); NASU (Ukraine); STFC (United Kingdom); DOE and NSF (USA).
Individuals have received support from the Marie-Curie programme and the European Research Council (European Union); the Leventis Foundation; the A. P. Sloan Foundation; the Alexander von Humboldt Foundation; the Belgian Federal Science Policy Office; the Fonds pour la Formation \`a la Recherche dans l'Industrie et dans l'Agriculture (FRIA-Belgium); the Agentschap voor Innovatie door Wetenschap en Technologie (IWT-Belgium); the Ministry of Education, Youth and Sports (MEYS) of Czech Republic; the Council of Science and Industrial Research, India; the Compagnia di San Paolo (Torino); and the HOMING PLUS programme of Foundation for Polish Science, cofinanced from European Union, Regional Development Fund.

\bibliography{auto_generated}   

\cleardoublepage \appendix\section{The CMS Collaboration \label{app:collab}}\begin{sloppypar}\hyphenpenalty=5000\widowpenalty=500\clubpenalty=5000\textbf{Yerevan Physics Institute,  Yerevan,  Armenia}\\*[0pt]
S.~Chatrchyan, V.~Khachatryan, A.M.~Sirunyan, A.~Tumasyan
\vskip\cmsinstskip
\textbf{Institut f\"{u}r Hochenergiephysik der OeAW,  Wien,  Austria}\\*[0pt]
W.~Adam, E.~Aguilo, T.~Bergauer, M.~Dragicevic, J.~Er\"{o}, C.~Fabjan\cmsAuthorMark{1}, M.~Friedl, R.~Fr\"{u}hwirth\cmsAuthorMark{1}, V.M.~Ghete, J.~Hammer, N.~H\"{o}rmann, J.~Hrubec, M.~Jeitler\cmsAuthorMark{1}, W.~Kiesenhofer, V.~Kn\"{u}nz, M.~Krammer\cmsAuthorMark{1}, I.~Kr\"{a}tschmer, D.~Liko, I.~Mikulec, M.~Pernicka$^{\textrm{\dag}}$, B.~Rahbaran, C.~Rohringer, H.~Rohringer, R.~Sch\"{o}fbeck, J.~Strauss, A.~Taurok, W.~Waltenberger, G.~Walzel, E.~Widl, C.-E.~Wulz\cmsAuthorMark{1}
% ...
\vskip\cmsinstskip
\dag:~Deceased\\
1:~~Also at Vienna University of Technology, Vienna, Austria\\
% ...

\end{sloppypar}
\end{document}